\documentclass[aps,pre,twocolumn,groupedaddress]{revtex4-1}
\usepackage{amsmath,amssymb,subfigure,fancybox,txfonts,bm,color}
\usepackage[dvipdfmx]{graphicx}
\begin{document} 
\title{
Atomistic Mechanisms of Intermittent Plasticity in Metals:\\
Dislocation Avalanches and Defect Cluster Pinning
}

\newcommand{\vc}{\mathbf}
\newcommand{\del}[2]{\frac{\partial #1}{\partial #2}}
\newcommand{\gvc}[1]{\mbox{\boldmath $#1$}}
\newcommand{\fracd}[2]{\frac{\displaystyle #1}{\displaystyle #2}}
\newcommand{\ave}[1]{\left< #1 \right>}
\newcommand{\intd}[1]{\text{d} {#1}}
\newcommand{\red}[1]{\textcolor[named]{Red}{#1}}
\newcommand{\blue}[1]{\textcolor[named]{Blue}{#1}}
\newcommand{\green}[1]{\textcolor[named]{Red}{#1}}
\newcommand{\eng}[1]{#1}

\newcommand{\secti}[1]{\section{#1}}
\newcommand{\jpn}[1]{}
\newcommand{\subti}[1]{}

\author{Tomoaki Niiyama}
\email{ni-yama@ike-dyn.ritsumei.ac.jp}
\author{Tomotsugu Shimokawa}

\affiliation{
College of Science and Engineering, Kanazawa University,
Kakuma-machi, Kanazawa, Ishikawa 920-1192, Japan
}
\date{\today}
\pacs{
}

\begin{abstract}

Intermittent plastic deformation in crystals with power-law behaviors 
has been reported in previous experimental studies.
The power-law behavior is reminiscent of self-organized criticality, 
and mesoscopic models have been proposed that describe 
this behavior in crystals.
In this letter, we show that intermittent plasticity 
in metals under tensile deformation can be observed
in molecular dynamics models, 
using embedded atom method potentials for Ni, Cu, and Al.
Power-law behaviors of stress drop and waiting time 
of plastic deformation events are observed. 
It is shown that power-law behavior is due to dislocation avalanche motions 
in Cu and Ni. 
A different mechanism of dislocation pinning is found in Al. 
These different stress relaxation mechanisms give 
different power-law exponents.
We propose a probabilistic model to describe the novel dislocation motion 
in Al, and analytically deduce the power-law behavior.

\end{abstract}

\maketitle

\secti{Introduction}
%

Many efforts have been devoted to understanding the plasticity of metals
from both fundamental and industrial motivations, 
but the nature of plasticity has not been fully understood yet.
Plastic deformation is a complex non-equilibrium process
with metastable structures, such as dislocations, disclinations, 
grain boundaries, or other defect structures.
Recent experimental and theoretical investigations into plastic deformation
from the viewpoint of non-equilibrium physics demonstrated 
intermittent and {\em power-law} behaviors,
$P(x) \propto x^{-\alpha}$,
in plasticity of crystals
\cite{AnanthakrishnaPRE1999crossover,AnanthakrishnaPhysRep2007,Miguel2001Intermittent-di,Richeton2006190,Dimiduk2006ScaleFreePlasticity,Csikor2007DislocationAvalanche,WeissPRB2007PowerLawExpMetal,Friedman2012NanoPillarSOC},
where $x$ is event size and 
$\alpha$ is a constant specifying the character of the system,
called a power-law exponent \cite{SOC1998Jensen,aschwanden2011SOC}.

Discontinuous deformation motions producing power-law behaviors 
in metallic alloys 
have been investigated experimentally 
by Ananthakrishna and his coworkers 
\cite{AnanthakrishnaPRE1999crossover,AnanthakrishnaPhysRep2007}.
Measurements of acoustic emission (AE) during creep of
ice and hexagonal-close-packed (hcp) single crystals having single-slip systems
showed acoustic energy bursts with power-law distributions
\cite{Miguel2001Intermittent-di,Richeton2006190}.
(Similar behaviors in AE measurements of metals had also been
reported in 1970s, the early years of AE techniques
\cite{Nakasa1976AE3rd,Nakasa1994AEBook}.)
In multi-slip crystals, such as face-centered-cubic (fcc) metals, 
the same behavior
was also observed in experiments and simulations
\cite{Dimiduk2006ScaleFreePlasticity,Csikor2007DislocationAvalanche,WeissPRB2007PowerLawExpMetal,Friedman2012NanoPillarSOC},
and it was pointed out that the existence of intermittency raises
problems for plastic forming of materials at micrometer scale
\cite{Csikor2007DislocationAvalanche}.
Studies of polycrystals
suggested the connection between the intermittency and 
mechanical properties of polycrystals (Hall-Petch's law)
\cite{Richeton2005Breakdown-of-av,Louchet2006HPLaw}.
In addition, similar intermittent behaviors were observed
in experiments and simulations of deformation in amorphous solids
\cite{Bailey2007amolSOCMD,Ng2009BMGserration,Sun2010MetallicGlassSOC,Budrikis2013AvalanchePRE,Salerno2012AmorphousAvalanches,Antonaglia2014BulkMetallicGlassSOC,Antonaglia2014BMGSOC}.
These studies indicate that such intermittent behaviors are 
a common property of plasticity, and understanding of such behaviors is 
crucial for revealing the nature of plasticity in crystalline materials, 
especially in metals.

The appearance of the power-law distribution suggests
the manifestation of \textit{self-organized criticality (SOC)},
tuned criticality, or other non-equilibrium critical phenomena
\cite{Bak1987SOC,SOC1998Jensen}.
SOC is known as a ubiquitous phenomenon in non-equilibrium systems,
and has been studied in many fields, including physics, geology, 
economics, and biology
\cite{SOC1998Jensen,aschwanden2011SOC}. 
However, it is not yet completely clear that the power-law behavior
in plasticity is related to SOC, tuned criticality, 
or other non-equilibrium critical phenomena.

There are some tentative explanations for the emergence
of intermittency and power-law behaviors in plasticity.
The {\em dislocation avalanches} description as one of the 
explanations for the intermittency
was suggested by discrete dislocation dynamics (DDD) simulations;
the intermittency arises from 
the self-organization of metastable jammed dislocation configurations
resulting from the long-range elastic interaction between dislocations
and avalanche-like motion of dislocations caused by
destruction of such metastable configurations
\cite{Miguel2001Intermittent-di,Csikor2007DislocationAvalanche}.
Elastic manifold depinning transition was also proposed as 
an explanation for intermittency
\cite{Zaiser2006IntermittentPlasticityReview}.
Thus, the mechanism of intermittency of crystalline plasticity
is still a controversial topic
\cite{Ispanovity2014NotDepinning}.

%

The dislocation avalanches were confirmed in not only single-slip crystals
but also multi-slip crystals, such as fcc metals, 
by three-dimensional DDD simulations 
\cite{Csikor2007DislocationAvalanche}.
However, unlike single-slip systems whose slips are dominated 
by only the elastic interaction between dislocations,
dislocation crossing (entanglements of lineal dislocations),
dislocation cross-slips, or other multi slip dynamics
might play an important role in multi-slip systems.
For instance, dislocation crossing can produce defect structures 
based on material properties, such as vacancy formation energy.
Such defect structures might play a key role in intermittent plasticity
due to the interaction between the defects and dislocation motion.
For complete treatment of such atomistic scale defect evolution
in multi-slip crystals,
atomic scale models are indispensable.

While many numerical simulations for intermittent crystalline plasticity
using mesoscopic scale models such as
a phase-field model, lattice spring models, cellular automaton models,
discrete dislocation models, a mean-field model,
and continuous models, etc. have been executed so far
\cite{Koslowski2004SOCphasefield,Salman2012SOCmass-spring,Budrikis2013AvalanchePRE,Csikor2007DislocationAvalanche,Miguel2001Intermittent-di,Richeton2005Breakdown-of-av,Dahmen2009plasticMFT,Papanikolaou2012Quasi-periodic,Ispanovity2014NotDepinning},
realistic atomic scale simulations, i.e., molecular dynamics (MD) simulations,
reproducing the power-law behavior have never been performed.

A numerical investigation similar to MD simulations 
has been performed 
by Moretti and his coworkers, who showed that
simulations employing two-dimensional overdamped dynamics
for small (colloidal) crystals
with a simple pairwise interparticle interaction 
reproduce power-law behaviors \cite{moretti2011yielding}.
Nevertheless, in order to study inherent features originating from atomic scale 
mechanical properties of individual materials,
more extensive simulations, 
such as three dimensional MD simulations 
employing realistic many-body interatomic potentials, are necessary.

In this article, we report that intermittent-plasticity with power-law behavior 
is exhibited in realistic three-dimensional atomic scale models of fcc metals 
with embedded atom method potentials
\cite{Mishin1999EAMAl,Mendelev2008EAMCu,Mendelev2012EAMNi}.
Specifically, MD simulation results are presented for nickel and copper 
that are consistent with power-law behaviors observed experimentally
\cite{Dimiduk2006ScaleFreePlasticity,WeissPRB2007PowerLawExpMetal}.
Results are also presented for aluminum whose 
vacancy formation energy 
and width of extended dislocations depending on the stacking fault energy 
are relatively smaller than those of Cu and Ni.
The simulation results show that the power-law behavior of Al is
different from that of Cu and Ni.
Based on these results we show how the material parameters affect
the non-equilibrium critical behaviors.

\secti{Model and Numerical method}
%

For the initial configuration of our MDs,
the perfect lattice configurations under periodic boundary conditions
described as follows were used.
At first, let us consider an fcc lattice with the lattice constant, $d$,
whose $x$-, $y$- and $z$-axis are along
the $(1 1 \bar{1})$, $(1 1 2)$, and $(1 \bar{1} 0)$ directions, respectively.
Next, we tilted the lattice around the $y$-axis with the angle 
$\theta = \tan^{-1} (3 \Delta L_z/2 \Delta L_x) \simeq 31.48$ degree,
where $\Delta L_x = \sqrt{33/2}d$, $\Delta L_y = \sqrt{3/2}d$, and
$\Delta L_z = \sqrt{11}d$.
This choice of orientation of the lattice is to reduce
slip planes for the sake of easiness of analysis
(the present orientation contains only two primary slip systems,
because these  have the largest Schmid factor
in this orientation
 \cite{Shimokawa2010GBMD}).
To fill a periodic boundary simulation cell with
this tilted lattice completely, we chose the cell dimension
 $( 8 \Delta L_x, 25 \Delta L_y, 5 \Delta L_z )$.
This cell is filled with $66000$ atoms and
will be a cubic shape whose length, $L$, is nearly $10$~nm
after 50 percent tensile deformation along the $z$-axis if the cell volume
remains constant.
In addition,
we also employed a larger cell with the dimension
$( 11 \Delta L_x, 35 \Delta L_y, 7 \Delta L_z )$
consisting of $177870$ atoms, which corresponds to a cubic dimension with 
$L=14$~nm after the deformation.

After relaxing the above configuration by the steepest 
descent method, we added random initial velocities to all atoms.
Next, we added uniaxial tensile deformation to the model along
the $z$-axis at the constant strain rate $\dot{\varepsilon} = 10^{9}$~1/sec 
during $500$ picoseconds (ps).
This strain rate is very much higher than that employed 
in experiments 
but it allows us to observe phenomena in the limited time period
that can be simulated with MD simulations.
Essential features of plasticity can be picked up
even under such a strain rate,
because the rate is sufficiently slow compared to
the velocity of dislocation motions during the MD simulations.
(The dislocation velocity in our MDs is comparable with 
the sound speed of crystals under the stress level.) 
This means that the external deformation works as a slow driving force
necessary for emergence of non-equilibrium critical behaviors,
especially SOC \cite{SOC1998Jensen}.
Moreover, it is known that some MD simulations
with extremely high strain rate conditions exhibit
typical mechanical properties, such as grain-boundary diffusion creep,
Hall-Petch relation and reverse Hall-Petch effect,
which agree with experimental results
\cite{Keblinski1998GBCreepMD,Schiotz2003reverseHP,Wolf2005GBMDreview}.

To suppress the thermal fluctuation concealing the stress drops due to
plastic deformation,
we controlled the temperature at $10$~K by velocity scaling.
The pressures along $x$ and $y$ directions were controlled at zero
by the Parinello-Rahman method without shear deformation \cite{Parrinello1980RahmanMethod}.
The MD simulations were realized by the velocity form of the Verlet algorithm
for integrating Newton's equation of motion,
with an integration time step $2$ femtoseconds.
Simulations were performed $20$ times with different initial velocities
of atoms.

\secti{Results and Discussion}

\begin{figure}[tbp]
\centering
  \includegraphics[width=7cm]{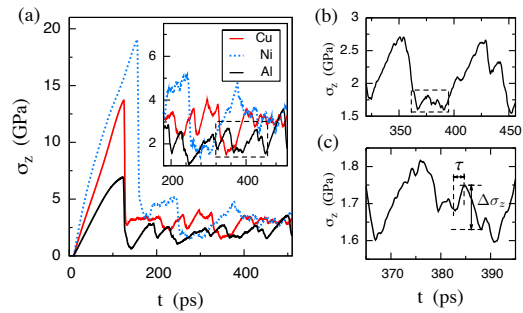}
  \caption{\footnotesize
    (a) (Color online) 
    Typical examples of the stress-time curve obtained from the MDs,
    where the curves of Cu, Ni, and Al are colored red (gray solid line), 
    blue (gray dotted line), and black,
    respectively.
    The curves in the range of $180$ - $510$ ps are imposed.
    (b) (c) The enlarged curves of Al.
  }
  \label{fig:t-stress}
\end{figure}

Typical examples of the resultant stress-time curves, $\sigma_z(t)$,
which were averaged
over $0.2$ ps interval to remove thermal fluctuations,
are depicted in Fig. \ref{fig:t-stress}(a).
After the first largest drops of the stress 
all the materials show serrated curves.
The noteworthy feature is that 
one can see self-similarity reminiscent of scale-free structure 
in the magnified serration curve
[Figs.~\ref{fig:t-stress}(b) and (c)].
This trend suggests that power-low behaviors occur in the deformation.

%

To examine statistics of these stress serrations,
we focused on stress drops originating from plastic deformation events
\cite{AnanthakrishnaPhysRep2007,AnanthakrishnaPRE1999crossover,Sun2010MetallicGlassSOC}.
Here we regard a time region when
the stress $\sigma_z(t)$ monotonically decreases
as the region of a plastic deformation event.
Thus, the value of stress drop is defined by the stress released
during one deformation event [Fig.~\ref{fig:t-stress}(c)]: 
$\Delta \sigma_z = \sigma_z(t_0+\delta) - \sigma_z(t_0)$,
where $\delta$ and $t_0$ are duration  and occurrence time of the event,
respectively.
Waiting time of each event is defined
by $\tau = t_0 - (t_0' + \delta') $, where $t_0'$ and $\delta'$ 
are the previous occurrence time and duration, respectively.
We calculated the stress drops and waiting times from
all the stress-time curves obtained from our MDs,
excluding the first stress drop events.

\begin{figure}[tbp]
\centering
  \includegraphics[width=7cm]{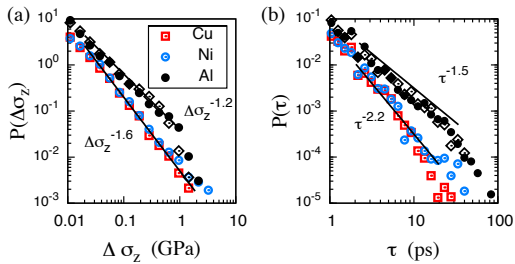}
  \caption{\footnotesize
    (Color online) Log-log plots of the probability distributions
    of (a) stress drop $\Delta \sigma_z$
    and (b) waiting time $\tau$.
    Open red squares, open blue circles, and solid black circles are 
    for Cu, Ni, and Al, respectively. 
    Open diamonds indicate the results  
    with the larger simulation cell for Al ($14$~nm on a side).
  }
  \label{fig:dist-stress-drop}
\end{figure}

The plots of probability distributions of stress drop and waiting time
in Fig.~\ref{fig:dist-stress-drop}
shows that all the distributions clearly follow power-laws 
with exponents $\beta_{s} \simeq 1.2$ - $1.6$ 
for stress drop and $\beta_{w} \simeq 1.5$ - $2.2$
for waiting time,
where the distributions of Al are shifted along the vertical axis
to facilitate visualization.
These results suggest that the plastic deformation of the present study
has strong intermittency and scale invariance for both time and event size.
The exponent $\beta_{s}$ is consistent with the previous studies
\cite{Miguel2001Intermittent-di,Dimiduk2006ScaleFreePlasticity,Csikor2007DislocationAvalanche,WeissPRB2007PowerLawExpMetal,AnanthakrishnaPRE1999crossover},
although precise evaluation of the exponents requires more extensive
studies because the values depend on the number of samples, 
averaging time interval  and event-detection techniques 
\cite{aschwanden2011SOC}.
Note that we confirmed that the exponents of the stress drop distribution
remain in the range from one to two
even if the averaging interval is made shorter.

Let us focus now on the apparent differences between the values
of the power-law exponents for Al compared to those of Cu and Ni.
It is well-known that the exponents reflect, in general,
the universality class to which the phenomenon belongs,
and the values depend on the universality class.
The intermittent plastic deformation of crystals is believed
to be in a certain common universality class 
\cite{Miguel2001Intermittent-di,Csikor2007DislocationAvalanche,Tsekenis2013UniClassDeterm}.
If this is true, one expects to obtain the same exponent values for
intermittent plastic deformation even if 
the material is  different.
However, the distributions of Al have smaller exponents,
($\beta_s \simeq 1.2$, $\beta_w \simeq 1.5$),
than those of Cu and Ni, ($\beta_s \simeq 1.6$, $\beta_w \simeq 2.2$)
[Fig.~\ref{fig:dist-stress-drop}].
(We also confirmed that this magnitude relation between 
the exponents of Al and Cu/Ni is independent of the average interval.)
This difference in the exponent values suggests that the deformation mechanism 
in Al is different to that of Cu and Ni, even though they are all fcc metals.

To clarify the deformation mechanisms in
Cu, Ni, and Al, we investigated the dynamics of defect atoms
during the deformation process by employing 
common neighbor analysis which can distinguish defect atoms
constructing dislocations, stacking faults, vacancies, and other defects from
lattice-structured atoms \cite{Jonsson1988IcoOrder,Clarke1993CNA}.
Typical snapshots of defect atoms in the present simulations,
obtained with an open visualization tool (OVITO) \cite{Stukowski2010Ovito},
are shown in Fig.~\ref{fig:snapshots} 
(for movies, see supplementary material \cite{supplyMovies}).
Figures \ref{fig:snapshots}(a) and (b) for Cu and Ni crystals
shows the formation of some stacking faults (colored pink or light gray), 
which span the periodic cell,
and interfacial dislocations on the stacking faults.
These dislocations have obviously a linear shape
[Fig.~\ref{fig:snapshots}(d) and (e)],
and showed intermittent collective behavior due to
the elastic interaction between them \cite{supplyMovies}.
This picture is consistent with dislocation avalanches 
\cite{Miguel2001Intermittent-di}.
In contrast with these results for Cu and Ni, 
the results for Al showed that, instead of stacking faults, 
a number of defect or disordered structures (colored blue or dark gray) 
were created
by nucleation, annihilation 
and crossing of dislocations [Fig.~\ref{fig:snapshots}(c) and (f)]
(for movies, see also supplementary material \cite{supplyMovies}).

\begin{figure}[tbp]
\centering
  \includegraphics[width=7cm]{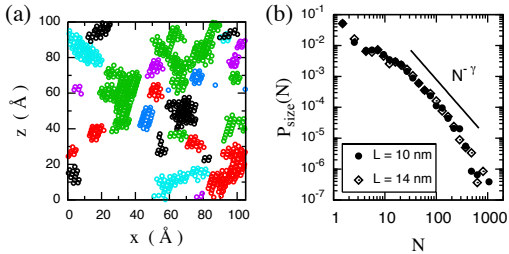}
  \caption{\footnotesize  
    (a) (Color online) Snapshot of defect atoms on a section diagonal
    to the simulation cell of Al, where projected coordinates on
    the $x-z$ plane are shown. 
    The atoms are colored according to their clusters.
    (b) Log-log plot of the size of defect clusters in Al.
    Solid circles and open diamonds are results for the simulation cells with
    side length $L=10$ and $14$~nm, respectively.
  }
  \label{fig:dist-N_def}
\end{figure}

The constituent atoms of these defect structures in Al multiplied and 
formed dozens of clusters during the tensile deformation.
We show a typical configuration of defect atoms in Al at $500$ ps 
on a sliced section diagonal to the simulation cell 
in Fig.~\ref{fig:dist-N_def}(a), where the section has $0.63$~nm thickness.
The size of these defect clusters has a wide range and,
the distribution seems to have a power-law tail
as shown in Fig.~\ref{fig:dist-N_def}(b):
\begin{align}
 P_{size}(N) \propto N^{-\gamma}, 
\label{eq:P_size}
\end{align}
where $N$ is the number of atoms constituting
one defect cluster.
While the precise estimation of the exponent $\gamma$ 
is difficult due to the curvature of the distribution originating
from its cut-off, 
one can roughly estimate $\gamma$ to be around $2$.

The MD results for Al show the behavior that
dislocations are often pinned by the clusters and then abruptly depinned, 
similar to the dislocation motion in crystals containing precipitates.
During the process, defect clusters spontaneously 
and slowly increase and decrease in size.
This implies that defect clusters act as quenched disorders
for dislocation motion.
It is expected that this difference between the dislocation behavior
of Al and that of the dislocation avalanche picture
is the reason why
different power-law exponents are obtained for Cu/Ni and Al.

These clear differences between Cu/Ni and Al 
are ascribable to material parameters such as
stacking fault energy and defect formation energy.
The lower stacking fault energy of Cu and Ni supports
the development of stacking faults in these materials.
The development of defect clusters which effectively change the relaxation
dynamics is attributed to relatively low vacancy formation energy of Al.
In other words, the values of material parameters such as
vacancy formation energy can affect the critical behavior in
crystalline plastic deformation.

The exponent of Cu and Ni, in which
domains are divided by twin boundaries,
is larger than the exponent of Al which is without twin boundaries.
This trend might be related to extremely high strength and ductility
in copper with high density nano twins \cite{Lu2004NanoTwinedCu},
because the larger exponent means suppression of large scale deformation
events.
Therefore, the investigation of power-law behaviors in the intermittent
plasticity may give us 
insights for designing nanostructure materials with excellent 
mechanical properties.

\begin{figure}[tbp]
\centering
  \includegraphics[width=7cm]{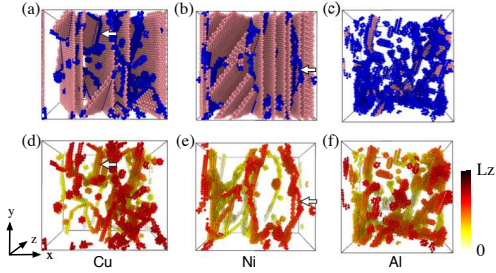}
  \caption{\footnotesize
    (Color online) Snapshots of defect atoms 
    in Cu at $272$ ps [(a) and (d)], Ni at $207$ ps [(b) and (e)],
    and Al at $211$ ps [(c) and (f)].
    The same configurations as the upper panels are shown in the lower panels, 
    where defect atoms are colored according to their $z$-coordinate,
    from $z = 0$ to maximum cell length $L_z$
    and stacking fault atoms are removed.
    The white arrows indicate one of the linear interfacial dislocations.
    (see Supplemental Material \cite{supplyMovies}.)
  }
  \label{fig:snapshots}
\end{figure}

%
\secti{Defect cluster pinning picture}
%

In order to explain that pinning and depinning motion 
in Al can produce power-law behaviors, 
we propose a basic model describing this type of dislocation motion
and prove that the model shows power-law behaviors.

As described above, a defect cluster in Al
works as an obstacle to dislocation motion.
The pinned dislocation would cut through the cluster
by an increased external shear stress $\sigma_{e}$ originating
from tensile deformation,
and then would be released from the pinning.
After that, the dislocation moves until being trapped by another cluster.
This movement corresponding to an individual deformation event
relaxes the external stress and produces a stress drop $\Delta \sigma$.

Here we propose a simple description of the above behavior in Al as a
one dimensional probabilistic model by employing the following assumptions.
First,
we exclude the formation and development of the defect clusters
from this model, and assume that the clusters already exist.
We assume that (i) 
there is only one straight dislocation 
which collides with the clusters one at a time.
(ii)
Migration of the dislocation from one cluster to the next
causes plastic deformation and decreases the external stress 
$\sigma_e$ by a constant value $\delta \sigma$.
(iii)
We here consider that a defect cluster serves as a precipitate in crystals.
It is well-known that the critical resolved shear stress of 
a precipitate is, in most cases, proportional to
some power of the size of the precipitate 
(e.g., see the section 5.6 in the reference 
\cite{Courtney2000MechBehaviorMaterials}).
Thus, the resistance, $\tau_R$, of the cluster can be described 
as a function of its size $N$: 
$\tau_R \propto N^{\eta}$, where $\eta$ is a constant.
(In general the resistance of a precipitate is represented 
by a function of its diameter.
However, introducing  an atomic volume 
one can easily transform the diameter to the number of
atoms composing the precipitate.)
Using the size distribution of a defect cluster [Eq.~(\ref{eq:P_size})],
the resistance is also determined by the power-law:
\begin{align}
P_R(\tau_R) = {{\tau_R}^{-\lambda -1}}/{Z}
\label{eq:P_R}
\end{align}
where $Z$ and $\lambda$ are a normalizing constant and power-law exponent,
respectively.
The exponent can be evaluated as 
$\lambda = (\gamma-1)/\eta$ by transforming 
the random variables in Eq.~(\ref{eq:P_size}).
Introducing the lower limit $\tau_{min}$ to avoid the divergence
of integration,
one can obtain $Z = \tau_{min}^{-\lambda}/\lambda$.
(iv)
If the external stress is less than 
the resistance of an encountered cluster
($\sigma_e < \tau_R$), the dislocation is trapped by the cluster
and the present deformation event is finished.
Subsequently, $\sigma_e$ increases due to the external tensile deformation,
then the trapped dislocation will be depinned when the increased 
$\sigma_e$ exceeds $\tau_R$ of the cluster.
On the contrary, (v) if $\sigma_e > \tau_R$,
\emph{the dislocation passes the cluster without being trapped
and migrates to the next cluster.}
This assumption allows the dislocation to pass multiple clusters
before being trapped and to produce a large stress drop.
When the dislocation passes $n-1$ clusters and is trapped by
the $n$-th cluster
the stress drop will be $\Delta \sigma = n \delta \sigma$.

From the above theoretical model,
we analytically deduce the probability $P(n)$ 
that one deformation event releases stress $n \delta \sigma$.
To evaluate $P(n)$ we consider 
the cumulative probability distribution $P_{cum}(n)$,
that the stress drop exceeds $n \delta \sigma$,
corresponding to the probability that a dislocation passes 
more than $n-1$ clusters without being trapped.
Here the probability that a dislocation passes a cluster
under the external stress $\sigma_e$, i.e. the probability that
an encountered cluster has resistance less than $\sigma_e$,
is calculated by the following: 
\begin{align}
G(\sigma_e) = \int^{\sigma_e}_{\tau_{min}} P_R(\tau_R) \intd{\tau_R} 
= 1 - \left(\sigma_e/\tau_{min} \right)^{-\lambda}.
\label{eq:Gs}
\end{align}
If a dislocation was pinned by a cluster with resistance $\tau_R$ at first,
$\sigma_e$ will be $\tau_R$ when the dislocation is depinned
from the cluster.
Therefore, the probability that the dislocation passes more than $n-1$ clusters
after depinning from the cluster is denoted by
$P_{cum}(n\ |\ \tau_R) = \prod_{k=1}^{n-1} G(\tau_R - k \delta \sigma)$.
Averaging this conditional probability over $\tau_R$ leads to
the cumulative probability:
\begin{align}
 P_{cum}(n) = \int^{\infty}_{n \delta \sigma} P_R(\tau_R)
\ P_{cum}(n\ |\ \tau_R) \ \intd{\tau_R},
\label{eq:P_cum}
\end{align}
where the lower limit $n \delta \sigma$ is introduced because of the condition,
$\tau_R - n \delta \sigma \ge 0$, for the positivity of $G(\tau_R - k \delta)$.

This integral is complicated so we adopt the following approximations.
Firstly, the term
$\left[ \left( {\tau_R-k \delta \sigma} \right)/{\tau_{min}} \right]^{-\lambda}$
can be considered small because of the smallness of $\tau_{min}$.
From this 
$\prod_{k=1}^{n-1} G(\tau_R - k \delta \sigma)$
can be approximated to only first order terms:
$
P_{cum}(n\ |\ \tau_R) \simeq 
1 - \sum^{n-1}_k \left( 
  \frac{\tau_R -k \delta \sigma}{\tau_{min}} \right)^{-\lambda}.
$
Secondly, noticing the smallness of $k \delta \sigma/\tau_R$,
one can expand and ignore the higher order terms;
$
\left( \frac{\tau_R - k \delta \sigma}{\tau_{min}} \right)^{-\lambda} 
\simeq {\left(  {\tau_R}/{\tau_{min}} \right) }^{-\lambda} 
(1 + k \lambda \delta \sigma/ \tau_R).
$
This approximation makes the summation computable,
so the integrand is rewritten in an integrable form at last:
\begin{align}
P_R(\tau_R)  P_{cum}(n\ |\ \tau_R) \simeq
{\tau_R}^{-\lambda -1} - \frac{n-1}{{\tau_{min}}^{-\lambda}}
{\left( {\tau_R}^{-1}  
+  \frac{ n \lambda \delta \sigma}{2} \right){\tau_R}^{-2\lambda -2} }.
\label{eq:integrand}
\end{align}
Now we can easily execute the integral. If we consider only the lowest term
of the result, one can obtain a power-law cumulative distribution:
$P_{cum}(n) \propto n^{-\lambda}$.
This result corresponds to the formula for the non-cumulative 
probability distribution,
$P(n) \propto n^{-\beta_{s}^{theo}}$,
where $\beta_{s}^{theo} = \lambda+1$ \cite{Sims2007JANE1208}.
This means that the present one-dimensional model has a power-law
behavior with the exponent $(\lambda+1)$.

Recalling the exponent, $\gamma \simeq 2$, obtained from 
Fig.~\ref{fig:dist-N_def}(b), 
we can now calculate the theoretical exponent of stress drop as 
$\beta_s^{theo} = \lambda + 1 = 1 + 1/\eta$.
Assuming $\eta = 5$, one can obtain the value $1.2$
which agrees with the value estimated in the MDs 
[Fig.~\ref{fig:dist-stress-drop}(a)].
This assumption lacks corroborative evidence,
but $\beta_s^{theo}$ remains around unity
unless $\eta$ is considerably smaller than unity.
This robustness suggests the essential correctness of the present model.

The present one-dimensional picture has some problematic assumptions:
Actually, a dislocation can come into contact with  multiple defect clusters
simultaneously.
Thus, our assumption that a dislocation encounters clusters sequentially
might be too simplified.
For the quantitative agreement of theoretical and numerical results,
an effective resistance distribution including the two-dimensional interaction 
between a dislocation and multiple defect clusters is required.
Deducing the actual distribution from numerical or theoretical
approaches is a subject for the future.
Also, the proportional relation between the resistance and size 
of the defective clusters should be established 
in future research, which might require some additional numerical simulations.
The most important aspect, however, is that
the difference between the theoretical and numerical exponents
is only small.
This agreement implies the adequacy of the present theoretical model
for depicting the intermittent plasticity in Al.

One might consider
the present model has some analogy to the model for domain-wall dynamics
in ferromagnetic materials, and the mean-field model 
which provides predictions that agree with 
some experimental results of intermittent plasticity
\cite{ABBM1990Barkhausen1,Dahmen2009plasticMFT,Friedman2012NanoPillarSOC}.
However, the former is a continuous model
and the latter is a cellular automaton model.
In contrast, our model is a probabilistic process.
The mean-field model and our suggested model
have obvious differences in the treatment of dislocation motion:
the present model describes the motion of dislocation itself,
as explained above.
Another significant difference
between them is
the distribution of threshold and arrest stress:
a polar function is employed in the mean-field model 
whereas the power-law distribution is used in our model.
This difference might explain the different values of the power-law exponents
originating from dislocation avalanches
and defect cluster pinning.

It is known that the power-law exponent of the avalanche distribution
of the domain-wall model for ferromagnetic materials
depends on the rate of the external field \cite{ABBM1990Barkhausen1}.
Thus, from the similarity between the domain-wall model
and our theoretical model, one might suppose that
the power-law exponent of the stress drop in Al also depends on
strain rate of tensile deformation.
However, the exponent is not expected to be as sensitive to the rate
as in the domain-wall model.
The reason is that experiments which employed strain rates more than
ten orders of magnitude slower than in our simulations
have obtained similar exponent values to the values obtained
from our MD simulations 
\cite{Dimiduk2006ScaleFreePlasticity,Richeton2006190,Friedman2012NanoPillarSOC}.
The clarification of the rate-dependence is an issue
for future study with more extensive MD simulations.

The approximations used in the above to reduce Eq.~(\ref{eq:integrand})
are equivalent to
ignoring the term $P_{cum}(n\ |\ \tau_R)$ in Eq.~(\ref{eq:P_cum}).
This means that the distribution of resistance supposed from the size
distribution of defect clusters,
$P_{size}(N)$, is the primary origin of power-law behaviors.
Thus, in this model
the scale invariant feature of self-organized defect
clusters produces the intermittent plasticity in Al.

                      \secti{Conclusion}

In order to investigate 
the intermittent crystalline plasticity
and influence of material characteristics upon 
the criticality of the plasticity,
we have performed  molecular dynamics simulations 
for uniaxial tensile deformation of Ni, Cu, and Al 
under the constant strain rate, temperature and pressure conditions
by using the embedded atom method potentials.
The simulations successfully reproduced power-law behaviors
of stress drop and waiting time of deformation events.
However, the power-law exponents for Cu/Ni and Al 
were found to be different.
Moreover, the dominant mechanism producing the power-law
behavior seem to be different in the case of Al.

Larger exponents ($\beta_s \simeq 1.6$)
are obtained from the power-law tails of the stress drop distributions 
of Cu and Ni,
and a smaller exponent ($\beta_s \simeq 1.2$) 
is obtained from the distribution of Al.
Dislocation avalanche like behaviors were observed in Cu and Ni,
whereas Al showed depinning motion of dislocations 
from self-organized defect clusters.

To describe how the latter behavior produces a power-law distribution
we proposed a simple one-dimensional model
and analytically deduced the power-law distribution from the model.
This model indicates that scale invariance of defect clusters produces 
the power-law features of the intermittent plasticity in Al.
The results indicate that differences in materials can affect
the universality of the critical behavior in intermittent plasticity
of crystals.
In order to confirm this, it is necessary to accurately determine 
the critical exponent values.
On the other hand, the difference of the power-law exponents in Al and Cu/Ni
might be an artifact depending on measurement method or conditions, such as
system size, boundary condition, strain rate, temperature, inertia etc.
To elucidate the above points more extensive studies employing 
molecular dynamics simulations are required.

\begin{acknowledgments}
We would like to thank K.~Dahmen and P.~Liaw for useful discussions and
comments.
This research was supported by the Ministry of Education, 
Culture, Sports, Science and Technology (MEXT) 
KAKENHI Grant Number 22102007 and the Japan Science and 
Technology Agency (JST) under Collaborative Research Based on 
Industrial Demand ``Heterogeneous Structure Control: 
Towards Innovative Development of Metallic Structural Materials''.
\end{acknowledgments}

\bibliography{/Users/niyama/Documents/Articles}

\end{document}